\documentclass[]{acm_sen_article}

\usepackage{lipsum}
\usepackage{flushend}

\usepackage{graphicx} 
\PassOptionsToPackage{hyphens}{url}
\usepackage{hyperref}
\usepackage{xspace}
\usepackage[numbers]{natbib}
\usepackage{balance}
\usepackage{xcolor}
\hypersetup{
  colorlinks=false,
  pdfborder={0 0 0}
}

\usepackage{fontawesome5}
\usepackage{framed}
\definecolor{formalshadelight}{RGB}{242,242,242}
\definecolor{formalshadedark}{RGB}{166,166,166}

\usepackage{csquotes}

\begin{document}

\title{The Human Need for Storytelling: Reflections on Qualitative Software Engineering Research With a Focus Group of Experts}

\subtitle{ACM SIGSOFT SEN-ESE Column}

\numberofauthors{2} 
\author{
Roberto Verdecchia\\
       \affaddr{University of Florence}\\
       \affaddr{Italy}\\
       \email{roberto.verdecchia@unifi.it}
\and
Justus Bogner\\
       \affaddr{Vrije Universiteit Amsterdam}\\
       \affaddr{The Netherlands}\\
       \email{j.bogner@vu.nl}
}

\maketitle

\begin{abstract}
From its first adoption in the late 80s, qualitative research has slowly but steadily made a name for itself in what was, and perhaps still is, the predominantly quantitative software engineering (SE) research landscape.
As part of our regular column on empirical software engineering (ACM SIGSOFT SEN-ESE), we reflect on the state of qualitative SE research with a focus group of experts.
Among other things, we discuss why qualitative SE research is important, how it evolved over time, common impediments faced while practicing it today, and what the future of qualitative SE research might look like.
Joining the conversation are Rashina Hoda (Monash University, Australia), Carolyn Seaman (University of Maryland, United States), and Klaas Stol (University College Cork,  Ireland).
The content of this paper is a faithful account of our conversation from October 25, 2025, which we moderated and edited for our column.
\end{abstract}

\section{Focus Group Discussion}
\textbf{Moderators:} \textbf{\textit{As a start to this focus group, we kindly ask everyone to briefly introduce themselves.
Who are you, and what's your relationship to qualitative research?}}

\textbf{RH:} I am Rashina Hoda, a professor of software engineering at Monash University in Melbourne, Australia.
My relationship with qualitative research is about two decades old.
Ever since my PhD, I've always loved it, occasionally been frustrated by it, but mostly been curious about it.
So, I continue to work on it. I still love it.

\textbf{CS:} My name is Carolyn Seaman.
I'm a professor of information systems at the University of Maryland, Baltimore County, in the United States.
I started my journey with qualitative methods also during my dissertation work in the last millennium, a very, very long time ago.
I found myself as one of the few, possibly only, SE researchers doing qualitative work early on, which was an interesting experience.
I do a lot of teaching of qualitative methods, both at my university and as an invited lecturer in many different places around the world.
So, I really like teaching it because it gives me the chance to reflect.
I also still conduct qualitative research and still coach my students through employing qualitative methods but also have a nice view of how it has evolved, in particular, the publication aspects of qualitative research.
Rashina, I, and other people also sat together and did a retrospective on qualitative research a year ago for the Transactions on Software Engineering journal (TSE)~\cite{seaman_qualitative_2025}.
So, I'm looking forward to reflecting and thinking back a bit in history for a second time.

\textbf{KS:} My name is Klaas Stol.
I'm a professor of software engineering at University College Cork in Ireland, and I'm also affiliated with Lero, which is the Research Ireland Centre for Software.
Well, where should I begin?
I guess I worked primarily qualitatively, but then I slowly shifted to quantitative methods.
In my PhD, I definitely was sometimes thrown a little bit into the \enquote{deep end}, and so I kind of felt like I taught myself qualitative analysis without having read too much initially.
And once you think that you've somewhat learned it, at some point I thought \enquote{I know how this works}. But then you read a little bit more, and then you realize \enquote{I know nothing}.
Early on, I was reading a lot of stuff on qualitative methods and grounded theory.
Today, Rashina will be the go-to person for that.
I remember her ICSE paper in 2010 well~\cite{hoda_organizing_2010}, which was one of the milestones for the community, just like Carolyn's 1999 TSE paper~\cite{seaman_qualitative_1999}.
Around that time, I was also observing many other papers falsely claiming to have done grounded theory, which bothered me a little.
So, that led to my 2016 ICSE review paper~\cite{stol_grounded_2016}, where I finally got to complain that \enquote{you're doing it all wrong!} (\textit{laughs})

\textbf{Moderators:} \textbf{\textit{Great, thank you all!
Then, let's dive in.
We want to start with a brief statement by everyone on why you think that qualitative research is important for SE.}}

\textbf{CS:} Oh, lots of different reasons!
I think the most important one is that software doesn't come into being by itself.
At least it hasn't up until now. (\textit{laughs})
Sometimes, it does now.
But anyway, my point is that there are humans involved.
And to study people and human behavior, which is what software development is for the most part, you need qualitative methods because humans are notoriously difficult to quantify.
You just miss a whole lot of information about people and how we do things by relying solely on quantitative methods.

\textbf{KS:} Absolutely, not everything can be done using quantitative methods.
It somehow limits the type of questions that you can answer.
As a very simple example, the standard t-test allows you to answer a certain type of question.
Of course, there's also a lot of variation in quantitative methods, so we can address more interesting and complicated questions.
But ultimately, they cannot answer questions that qualitative methods can, in particular, exploration and gaining an understanding of ideas by simply observing and talking to people rather than only guessing what might be a particular mechanism based on some evidence via a quantitative method.
In short, going to the source rather than trying to guess things.

\textbf{RH:} I definitely agree with all of those points, but I also think that the power of storytelling is something very central to qualitative research.
One can argue that you can tell stories with quantitative data, but go right back to cave people times and you find stories on cave walls.
There's just an inherent need for people to capture and tell stories, and I think that kind of expands into research.
We need to tell our stories of what's going on in practice, and I think nothing quite does the job of storytelling like qualitative research.
So, I'm a big fan of this aspect, especially bringing in the raw data as much as possible in terms of, e.g., quotations from interviews or photographs of observations.
It's this window into the actual lived experience that you just don't get with numbers.
And as Klaas said, there are other things for which numbers are very important, so I don't think it should be a \enquote{qual vs. quant fight}, although it oftentimes comes across as such.
But for me, it's mainly about the very inherent human need to tell stories.

\textbf{Moderators:} \textbf{\textit{Very clear, thank you! Now that we've established why we need qualitative SE research, we want to pick your brains about how it was at the beginning of your academic career: if you look back 10 to 20 years, how was your experience with publishing qualitative studies? And also, how has this changed over time?}}

\textbf{CS:} So, when I was trying to put together my PhD thesis proposal, I wanted to study something that was quite human-focused, which was already a little unusual at the time.
My dissertation advisor was a very quantitative, very empirical person.
But I was really struggling with quantifying the things that I wanted to study.
One day, my advisor came back from serving on a committee for someone in a business school, somewhere in upstate New York, and he said: \enquote{There's this thing called ‘qual something'. You might want to look it up.}
So, I did.
There was, of course, nothing for SE at the time, so I was looking at stuff from the educational and social sciences.
It was super thick and dense, and I didn't understand most of it, but I just started trying it.
So, I was very much self-taught, like Klaas also mentioned.
I tried to take courses in the education and psychology departments, but they wouldn't take me seriously because I was from the computer science department, so they wouldn't let me in.

As time went on, I was able to publish the mentioned 1999 paper in TSE~\cite{seaman_qualitative_1999} and also work from my dissertation, which led to me becoming the \enquote{SE qual person} for a while.
Then, other people started getting on the bandwagon, which made publication very difficult because there were very few people who could actually review such papers.
I was getting lots and lots of review requests, but I couldn't review everything.
It was tough for a while because people just didn't know what to do with qual papers.
So, I started advising people that they should do mixed methods.
That's what I had to do with my dissertation: I had to put numbers in.
Otherwise, I wouldn't have a degree from the computer science department today.
I think, for a long time, mixed methods was the \enquote{shield from the publication woes} for many people.
That was a way to get stuff published but also to sneak in some qualitative work.
I think that also helped a lot to get people more used to it.
Eventually, more people came on board, so more people were able to review.
However, early on, I think a lot of really bad qualitative work was published, while a lot of good qualitative work wasn't.
So, \enquote{published or not} was a very crude metric for a long time.

It's a different world now, and there are many qualitative SE researchers.
For me, the world has also opened up in how many different ways you can do qualitative research.
I learn about new methods all the time, which is really exciting because – guess what? – there are other people who do research in the world who aren't interested in SE or even tech topics.
And they use all of these cool methods.
I think the SE community can actually pat itself on the back a little bit because we've done a good job of borrowing things from other fields and making them our own.
However, there's lots more out there to explore.

\textbf{RH:} I was smiling and nodding along because of very similar experiences, although I started much later than Carolyn.
But I am definitely also familiar with experiences like isolation, confusion, or being the weirdo in the group who is talking to people instead of running some algorithms and numbers.
It was very isolating at the beginning.
I had a similar journey in terms of trying to figure it out on my own, trying to make sense of books that were written by sociologists for sociologists.
As a software engineer, I asked myself, where is the flowchart? What are the steps? What am I supposed to do first? Can I get this all on one page? It was extremely frustrating to understand all of that from this very narrative style of reading across a number of books.
I was doing grounded theory, and it was really hard in terms of publishing this.
Almost all my papers back in the day felt like a fight to get them in.
But I also made it especially hard for myself because I was actually focusing on the top venues.
I said to myself, \enquote{We're doing good work. Why does this not get into the top venues? I need this in the top venues.}
So, I made it actually harder for myself.
In the end, it was also satisfying to get there, but it was a fight.
It didn't come easy.

I still remember one of my response letters, which was way longer than the actual paper.
I learned so much because I literally was just defending myself all the time, like a defense before the PhD defense.
But that actually ended up being one of my most cited papers.
Many of my memories are filled with challenges and frustrations, but also an incredible amount of learning and resilience.
I've also been able to pass on a lot of this learning to students and colleagues, so it definitely wasn't all negative.
I think the thing that really kept me going was the mentioned storytelling.
When I would go into industry for my interviews and observations and actually talk to the practitioners, it was pure bliss.
Just listening to their stories of how something actually works and why or when something doesn't work: \textit{that's} what kept me going.
I also went back and presented my findings, not just to academia, but also to practitioner communities like meetups and agile conferences.
The comments I got from practitioners were really validating because I would see how the stories resonated with them.
In these moments, I knew we were doing something right even though reviewer 2 didn't always agree with us.
The practitioners could see \textit{their} stories in \textit{our} stories.

\textbf{KS:} I was trained 
at Lero, and there were other people doing interviews as a research method, so I didn't feel isolated.
I just felt that I didn't really know what to do.
So, my supervisor would send me papers to read, basically implying that I would just pick up whatever skills those authors had.
But I really learned a lot from that.
One that stands out is a 2008 ICSE paper by Cleidson de Souza and David Redmiles~\cite{de_souza_empirical_2008}, when qualitative methods were still relatively scarce.
If you browse that paper, it only has one picture, and the rest is just text.
It's just walls of text.
Normally, this sounds very negative, but I don't mean it like that.
One thing I took away from that paper was how they structured and organized it.
It's all very well that you do interviews, but then you have all these transcripts, and you try to make some sense of them, e.g., what is it really that you need to look for?
In the paper, they had this simple figure with only about four entities and three arrows.
They organized the structure of the paper along that figure.
It's so simple, like primary school stuff, but that kind of made the penny drop for me.
You declare this initial framework and say \enquote{I think the world looks like this}, and then you talk about all those elements and how they unfold in real life.
I learned how to do this from that paper.
That's why I'm also a huge fan of research frameworks.

The other pivotal moment for me was when an information systems (IS) researcher, namely Matt Germonprez, came to visit Lero.
In IS, they have a very different way of doing research.
In SE, we usually recruit a bunch of software developers and just capture their challenges and all those things.
Everything is about challenges.
IS researchers don't think like that.
They take a very different approach where they are very focused on a \textit{theory}.
And then they really treat the actual thing they study and the actual people they talk to as a vehicle to advance that theory.
Back then, Matt presented his 2017 ISR paper about a theory of responsive design for corporate engagement with open-source communities~\cite{germonprez_theory_2017}.
That really made me change the way I looked at things and why we do qualitative research rather than just reporting what you found.
My first attempt to publish like this was a case study.
I had read the Robert Yin book about case studies~\cite{yin_case_2018} and found it very sterile.
The first time you read this book, you kind of get it but not really.
Yin talks about propositions and all that, which even sounds a little bit quantitative.
And then you go into the field and wonder how you can make this work.
I recently reread the Yin book and finally understood what he meant.
My case study was obviously rejected at ICSE due to reviewers complaining about a lack of generalizability and what we could learn from this, i.e., the typical comment of that era.

\textbf{Moderators:} \textbf{\textit{Let's continue with a question about your favorite success story of a qualitative study of yours, e.g., one that you particularly liked or that you think was very successful or impactful.}}

\textbf{RH:} Oh, I've got quite a few.
A lot of my students' stories these days are my favorites.
Recently, Hashini Gunatilake finished her PhD on the role of empathy in software engineering.
She has done a fabulous job in the number of different ways in which she investigated the topic.
One of them is this theory on the role of empathy in software engineering~\cite{gunatilake_empathy_2025}, which was published this year in TOSEM.
The theory is presented using the 6Cs template, and then – as Klaas mentioned for the other paper~\cite{de_souza_empirical_2008} – the rest describes the theory via the different parts of the diagram.
So, it becomes like a signpost for the paper, from which the structure follows quite easily and where you can embed the quotations and so on.

\textbf{KS:} My chosen paper is an ICSE 2014 paper, a case study of crowdsourcing software development~\cite{stol_crowdsourcing_2014}.
I like this paper very much, as it's my first success story.
There were papers before that, but I really owned this one.
We were flying back from ICSE 2013 and had just presented a paper there.
Brian had showed me the ropes on how to do this, like not explicitly, but he just went through the process, and I was observing and absorbing everything and thought, OK, I can do this.
Before my trip to ICSE 2013, I had downloaded like 100+ papers on crowdsourcing from different fields.
On the flight back, I remember I just went through them one by one very quickly and started writing down the keywords, like what are the main themes.
That's how I developed the six-element framework that ultimately became the ICSE 2014 paper.
I realized that all these papers are talking about a bunch of things like coordination, planning, remuneration, or motivation.
There were six themes that kept reoccurring and that I synthesized into a framework.
There was also a part about knowledge management and intellectual property.
I remember Brian pushing back against this, but after I explained it to him, he agreed to keep it.
That was the moment that I felt like I achieved something and that I started to approach his level of thinking.
I still think it's a nice paper in hindsight, but it also feels a bit too simplistic to me today.
Nonetheless, this was my first big personal success.

\textbf{CS:} I decided on an interview study that was published in 2007 in the IST journal~\cite{lutters_warstories_2007}.
It is about the value of war stories and debunking myths about documentation and software maintenance, which was a sort of precursor to my interest in technical debt.
I was collaborating with my friend Wayne Lutters, who comes from the field of computer-supported cooperative work and who introduced me to the war stories technique.
It is an interviewing technique where you ask people for specific stories with a certain structure and certain attributes that make something a war story.
It was basically a storytelling type of interview structure, which I really enjoyed.
We had a lot of fun just asking people to tell stories, and some people really got into it.
Some of our interviewees were just master storytellers!
But what was fun too was that this method really took hold at several conferences for a number of years afterward.
Many people came up to me and said, \enquote{Oh, that war stories thing, that sounds like fun.
I'm going to try that!} And there were several other war stories papers that followed.
So, it was a very fun study to do because of that framework of viewing things as war stories.
The paper was pretty good too, but the methodology was particularly interesting.

\textbf{Moderators:} \textbf{\textit{What is, in your opinion, the relationship between qualitative studies and industry?}}

\textbf{CS:} In the early days, the validation came from practice.
This is a powerful relationship, which mirrors what happened way back when empirical work was trying to find its validation in computer science research.
Empiricism was really viewed with severe suspicion early on in the 80s and early 90s.
\enquote{\textit{That’s social science stuff, that's silly! You have to prove something! You need to have an algorithm!}}
What really validated empirical work back then were practitioners, who wanted to understand their processes and improve them.
And that translated into support from funding agencies.
So, empirical work started getting funding and then all of a sudden it became respectable.
At some level, the same thing happened with qualitative research.
I think action research did an important job in supporting qualitative research because nobody thought practitioners would be interested at all in participating in a research process.
Quantitative folks often just swoop in, pick up a bunch of data, swoop out, and then write a paper.
Qualitative researchers can't do that.
It takes a while.
Practitioners really appreciated this effort, which in the end led to more respect for qualitative work.

\textbf{RH:} I also think that qualitative research and industry are in a direct relationship.
Most of my work is industry-based anyway.
In the early days, it was about understanding some of the buzzwords that were thrown around agile software development, e.g., what terms such as self-organizing teams,  Scrum master, and product owner actually mean.
It was a minefield and a great opportunity to deep dive into each buzzword to really understand what it actually meant in practice.
What was very exciting then, and actually still is exciting now, is to hear those lived experiences and those stories of \enquote{here is what the book says} and \enquote{here is what we are actually doing}.
And the reason why the two can be different is also very important.
That's where qualitative research shines: grabbing onto that \textit{why}.

As Carolyn hinted at, it takes a bit of time to build the trust and to get to a point where practitioners are not simply regurgitating the seminal books.
In a lot of my initial interviews, people would start to literally rattle off the agile manifesto, and I had to explain that this is not that type of research.
I am not here to audit.
I'm not here to judge.
I want to know your real struggles, and I can guarantee full confidentiality and privacy.
That is a reputation that you build over the years to a point where people are forthcoming.
The level of honesty improves over time, to the point practitioners felt they could wash their dirty linens with me.
Confidentiality and trust is why it is so difficult to share raw data of qualitative research.
To counter that, you can provide plenty of anonymized examples, but you can almost never share all the raw data because that's a promise you made to your participants.
It is almost sacred.

\textbf{KS:} I think the relationship between qualitative research and industry can be many different things, but a meta-question underpinning all qualitative research is to understand \enquote{\textit{What's going on here?}}.
There is a study by Bill Curtis published in the Communications of the ACM in 1988, \enquote{A field study of the software design process for large systems}~\cite{curtis1988field}.
It is one of the first well-known qualitative studies appearing in a high-quality venue.
I'm still surprised today that it has not been picked up more.
It's a fantastic study that documents so many concepts it could have been published today.
It’s sad to reflect that we have not really learned more from it as a community due to the lack of cumulative tradition we have in our field.
The work by Curtis and others is a beautiful description of \enquote{what is going on}.
And based on personal experience, companies agree to partake in this kind of study because we provide them a mirror, showing them what is happening by documenting and reflecting on what they are doing.

Just to make another link to information systems: they take a slightly different approach from ours.
In software engineering, typically 20+ interviews would be enough to publish a journal paper in the Journal of Systems and Software (JSS) or even TSE.
In information systems, that would instead only be the first step to get a set of initial exploratory themes that need to be better understood through further research.
Qualitative research in software engineering has made huge steps in the past few decades, but there is still a lot that needs to be learned.

\textbf{Moderators}: \textbf{\textit{Thinking about the junior SE researchers that use qualitative research methods, what do you think are the most crucial impediments that they might face today?}}

\textbf{CS:} If you tell students in 2025 to find some good examples of qualitative software engineering research, it is hard to pick out the gems.
There are still some bad examples  that students might not be able to recognize, as these articles got cited and held up because people followed the same suboptimal practices.
Last month, I spent two weeks in Finland teaching qualitative methods.
It gave me the opportunity to see just how surprised students were about how involved qualitative methods actually are.
I started by talking about grounded theory, and they immediately wanted to use it.
But in reality, you should not do a grounded theory as your first qualitative study because it involves a huge amount of effort, and it's hard.
A pitfall lies in thinking that reading the literature is not necessary, that simply picking a research method does not imply first reading the literature.
There is still this impression that qualitative methods are easier than quantitative methods.
I'm not sure where that comes from.
Perhaps, bear with me, this is linked to page limits for papers.
It is fairly easy to describe a methodology succinctly, but not in a way that really shows all that went into it.
When I pick out good examples for students, I have to go for venues without page limits that allow for a thorough documentation of the methodology, and there are not many out there.
There are also a number of papers that do not report data analysis in their methodology section, which skip directly from data collection to results.
All these pitfalls perpetuate the misconception that qualitative research consists of merely talking to people and then writing a  paper about it.
Qualitative research might be fun, but it is not easier than quantitative research.

\textbf{KS:} I had a thought related to the mentioned page limits.
From a reviewer's perspective, I want to know everything about a research method.
However, as a writer, I cannot document everything I did.
Qualitative research is not a linear process.
It could go in all sorts of spirals and dead ends, but you cannot take a reader down all those paths, so we have to make it sound as if it was just one.
As a reviewer, I understand the frustration of not knowing all research details, but as an author, I think they should just trust me, which is obviously a terrible thing to say.
I take reviewer comments very seriously and try to do my best with them, but it is very challenging to precisely explain your thinking and mental model the way you are \textit{really} doing things.

\textbf{RH:} I was smiling the whole way through.
The points that both of you raised are so true.
I think the biggest challenges of qualitative research are -- funnily enough -- \enquote{quantitative}.
Whether it's not having enough length to explain your method and your findings in detail, with the quotations and raw data examples, or how much data is enough data.
These are quantitative issues.
And then, a whole lot of other things around generalizability and so on.
But how much data is enough is something that I am sure you guys also get asked at every single tutorial and technical briefing.
People just want a number, but it is not that simple.
However, I am also not a strong believer in \enquote{it depends}.
Of course it depends, but there are some figures that are usually accepted in the SE community because I do not want people to simply use an \enquote{it depends} answer and then they receive five rejections in a row before they find out for themselves.
I give some concrete advice about data quality, quantity, and theoretical saturation in Chapter 7 of my book on qualitative research~\cite{hoda_book_2024}.

\textbf{KS:} We can expand this to the more general notion of the \enquote{guidelinification} of everything.
Quantitative methods follow a bit of a more straightforward process, even if the statistics are not as objective as some people think.
You collect the data following a predefined process, analyze it with a particular technique, and draw your conclusions.
Qualitative methods are quite different.
Sometimes, you have to try out different things.
So, people need guidelines to do that.
As a community, we are pretty good at defining guidelines.
But here comes the paradox: people use the guidelines without taking the time to understand their actual \textit{meaning}.
They just follow them.
People just want to be told what to do, but the whole point is to figure out by yourself how qualitative research works.
The paradox is that, on the one hand, we want to help everybody and make things better by giving back as best we can by writing guidelines.
I am as guilty as anybody with my mentioned review paper~\cite{stol_grounded_2016}, which even has \enquote{guidelines} in the title.
However, on the other hand, trying to capture complex concepts in 10 bullet points, or even 300 pages, is really hard.
Even if you succeed, some people just follow guidelines to the letter of the law rather than in the spirit of the law.
Guidelines are just used as checkboxes to follow by some, e.g., step one, two, and three, without thinking about what the guidelines really try to convey, namely the ideas behind them.

\textbf{RH:} I cannot agree more.
About the joy of doing qualitative research, I feel sad for some of the newer generation.
There is so much pressure to publish.
You can see it both in papers and sometimes in presentations.
For example, somebody presents a mixed-method study where four different methods were used.
But as soon as I start questioning any one of them, you can instantly see the lack of depth.
They cited this reference for this method, another reference for that other method, and then they combined it all together and thought they are done.
But the lack of depth is just shocking sometimes.

\textbf{Moderators:} \textbf{\textit{Looking at the bigger picture, where do you see the future of qualitative research going? And what can we do to make sure it goes in the right direction?}}

\textbf{RH:} What's the two-letter word these days? (\textit{laughs}) \enquote{AI}?
Is anyone saying that?
Do we want to go there?
I don't know if we want to go there\dots

\textbf{KS:} Well, that's the elephant in the room.
Because researchers just upload their transcripts and ask for a thematic analysis of it.
And the problem with doing so is that you are not getting real insights: you are just getting a summary of the transcripts at best.
You are not gaining knowledge through learning, you miss the whole point.
Those who have done qualitative studies can talk about everything they learned.
We remember anecdotes, people, and experiences.
We can tell a cohesive story about the essence of a qualitative study.
If you use an LLM to do that, all you have left is a bland summary.
This is a huge problem.
Due to time pressure, attempting to submit a paper to a conference, maybe going to Hawaii, it is tempting to do so because it is a very low barrier.
It seems so enticing.

\textbf{RH:} AI is getting a lot better at handling things.
But what Klaas just mentioned about learning is also very related to reflexivity.
The journey of any research has two outputs: one is the study and the other is becoming a new self.
We, as qualitative researchers, are the filter through which all data passes.
We are going to have a profound impact on the stories we tell about what we have studied.
All of that is just stripped away when you use an LLM.
You can try to make LLMs do role-playing, e.g., to analyze a transcript like Carolyn does, but it is not the same thing.
Your ability to connect across time and space, to connect across languages and cultures: the human capacity to do that is just unbelievable.
Take Klaas's study on crowdsourcing~\cite{stol_crowdsourcing_2014} or Carolyn's study on documentation~\cite{lutters_warstories_2007}, their essence would not be the same if somebody else did them.
We cannot ignore the ability of humans to act like filters.
And unless we find a very reliable way to pass all the ability of humans to act like filters to LLMs, I am not going to be convinced.

\textbf{CS:} I have not played with the use of AI and qualitative methods at all, but let me play devil's advocate for a minute.
I do not think that practitioners are interested in people acting like filters, e.g., \enquote{Carolyn's view of documentation}.
They want something they can use.
AI could be used to represent different researchers, filters with different biases, to see what their convergence is.
I think that would be fascinating.
I have not seen anybody do it, and I certainly have not tried, but playing with this idea of researcher bias is interesting.
Right now, we only have human brains to play with this.
I always tell my students that qualitative analysis is like swimming: you should never do it by yourself.
Nowadays, we achieve this with collaborations, i.e., multiple brains.
I think there is real value in doing that.
But what if you scaled up this concept a million times with AI?
What would happen?
I don't know.
I struggle with the idea of researcher bias because it humanizes the results, makes us advocates of our work, and injects passion and human insight into the stories we tell, which I think is really valuable.
But sometimes, to make research have impact, it needs not to be personalized by a researcher.
There is a balance there, a dichotomy.
I think this is very interesting and that it would be great to play with this idea a bit.

\textbf{RS:} I think that is very fascinating.
I have played with LLMs for qualitative research, and obviously, at the surface level, they got better and better over time.
I would say it gets very tricky when we think about the future generations of qualitative researchers we were talking about.
That is what I worry about the most: LLMs in the hands of people who are not interested in knowing how to do qualitative research.
LLMs can confuse the living daylights out of them.
While LLM providers have obviously worked on this, they are still tools tailored to please and agree with you.
At the heart of it, I feel that somehow the \enquote{average-ification} (if there is such a word) that LLMs do is really conflicting with the whole idea of qualitative research, i.e., LLMs trying to be the average human.
There is no average human.
But I think this is the beginning of another great conversation.
We have just opened Pandora's box\dots

\textbf{Moderators:} \textbf{\textit{Thank you all for this interesting discussion!}}

\section{Conclusion}
It was difficult for us to reduce the rich and deep conversations we observed into a more concise form for readers that stays true to the original spirit of the discussion.
We therefore decided to report the focus group conversation in a fairly verbatim manner but could not incorporate everything for space reasons.
We hope that readers can still easily identify the parts that resonate with them the most and that the discussion can spark their curiosity or expand their way of thinking.
As conclusions, we also tried to pick some favorite takeaways for ourselves.

One of them is Rashina Hoda's point about qualitative studies having two outputs: \textbf{the study itself and a more knowledgeable version of the researchers.}
Observing, documenting, explicitly discussing and organizing concepts, and finally writing about them slowly converts initial personal intuition into a holistic bigger picture.
In this sense, qualitative studies in SE are a fairly unique opportunity for building expertise.
Another noteworthy highlight for us was the point about the importance of qualitative methods for SE research.
Klaas Stol clearly described the power of qualitative research for \textbf{exploration and
gaining an understanding of ideas}, e.g., by observing and talking to people, something that quantitative methods simply cannot do.
Similarly, Rashina Hoda connected qualitative research to the \textbf{inherent human need for storytelling}, while Carolyn Seaman highlighted the \textbf{industry interest in qualitative studies}, which played an important role in making qualitative research more mainstream.

Smaller but noteworthy points were the potential \textbf{negative impact of the page limits} of most SE venues on qualitative research (mentioned by all three experts) and the ability to \textbf{build trust with your participants as a key skill} of qualitative researchers (Rashina Hoda).
While these points may be familiar to more experienced researchers, they are not as often discussed in the respective methodological literature for qualitative SE research.
Lastly, while the \textbf{usage of GenAI for qualitative analysis} is still seen as fairly controversial, risky, and challenging by all three experts, they definitely acknowledged potential opportunities in this area.
How to incorporate GenAI into qualitative research responsibly is still an open question, and it's unclear whether the gained expertise and insights of manual qualitative analysis can ever be fully compensated for with potential efficiency and scalability gains.
After all, a core conclusion echoed by all participants was that, while studying the methodological literature is still important, \textbf{the best way to build deep expertise with qualitative methods is by applying them}, with potential reflections afterward.
So, let's go find some participants!

\section*{Acknowledgements}
We express our gratitude to Rashina Hoda, Carolyn Seaman, and Klaas Stol for their availability, passion, and priceless insights, which ultimately made this piece possible.

\bibliographystyle{plainnat}
\bibliography{biblio}

@article{curtis1988field,
title={A field study of the software design process for large systems},
author={Curtis, Bill and Krasner, Herb and Iscoe, Neil},
journal={Communications of the ACM},
volume={31},
number={11},
pages={1268--1287},
year={1988},
publisher={ACM New York, NY, USA}
}

@book{hoda_book_2024,
    author = {Rashina Hoda},
    title = {Qualitative Research with Socio-Technical Grounded Theory},
    publisher = {Springer},
    doi = {https://doi.org/10.1007/978-3-031-60533-8},
    year = {2024}
}

@inproceedings{hoda_organizing_2010,
    address = {Cape Town South Africa},
    title = {Organizing \textit{self-organizing} teams},
    isbn = {978-1-60558-719-6},
    doi = {10.1145/1806799.1806843},
    language = {en},
    urldate = {2025-11-18},
    booktitle = {Proceedings of the 32nd {ACM}/{IEEE} {International} {Conference} on {Software} {Engineering} - {Volume} 1},
    publisher = {ACM},
    author = {Hoda, Rashina and Noble, James and Marshall, Stuart},
    month = may,
    year = {2010},
    pages = {285--294},
}

@article{seaman_qualitative_1999,
    title = {Qualitative methods in empirical studies of software engineering},
    volume = {25},
    issn = {00985589},
    doi = {10.1109/32.799955},
    number = {4},
    urldate = {2022-12-13},
    journal = {IEEE Transactions on Software Engineering},
    author = {Seaman, C.B.},
    month = jul,
    year = {1999},
    pages = {557--572},
}

@inproceedings{stol_grounded_2016,
    address = {New York, New York, USA},
    title = {Grounded theory in software engineering research},
    isbn = {978-1-4503-3900-1},
    doi = {10.1145/2884781.2884833},
    booktitle = {Proceedings of the 38th {International} {Conference} on {Software} {Engineering} - {ICSE} '16},
    publisher = {ACM Press},
    author = {Stol, Klaas-Jan and Ralph, Paul and Fitzgerald, Brian},
    year = {2016},
    pages = {120--131},
}

@inproceedings{de_souza_empirical_2008,
    address = {Leipzig, Germany},
    title = {An empirical study of software developers' management of dependencies and changes},
    copyright = {https://www.acm.org/publications/policies/copyright\_policy\#Background},
    isbn = {978-1-60558-079-1},
    doi = {10.1145/1368088.1368122},
    language = {en},
    urldate = {2025-11-18},
    booktitle = {Proceedings of the 13th international conference on {Software} engineering  - {ICSE} '08},
    publisher = {ACM Press},
    author = {De Souza, Cleidson R. B. and Redmiles, David F.},
    year = {2008},
    pages = {241},
}

@article{germonprez_theory_2017,
    title = {A {Theory} of {Responsive} {Design}: {A} {Field} {Study} of {Corporate} {Engagement} with {Open} {Source} {Communities}},
    volume = {28},
    issn = {1047-7047, 1526-5536},
    shorttitle = {A {Theory} of {Responsive} {Design}},
    doi = {10.1287/isre.2016.0662},
    language = {en},
    number = {1},
    urldate = {2025-11-18},
    journal = {Information Systems Research},
    author = {Germonprez, Matt and Kendall, Julie E. and Kendall, Kenneth E. and Mathiassen, Lars and Young, Brett and Warner, Brian},
    month = mar,
    year = {2017},
    pages = {64--83},
}

@book{yin_case_2018,
  title={{Case Study Research and Applications: Design and Methods}},
  isbn = {978-1-5063-3618-3},
  author={Yin, Robert K.},
  year={2018},
  publisher={SAGE Publications}
}

@article{gunatilake_empathy_2025,
author = {Gunatilake, Hashini and Grundy, John and Hoda, Rashina and Mueller, Ingo},
title = {The Role of Empathy in Software Engineering - A Socio-Technical Grounded Theory},
year = {2025},
publisher = {Association for Computing Machinery},
address = {New York, NY, USA},
issn = {1049-331X},
doi = {10.1145/3768315},
journal = {ACM Trans. Softw. Eng. Methodol.},
month = sep
}

@inproceedings{stol_crowdsourcing_2014,
author = {Stol, Klaas-Jan and Fitzgerald, Brian},
title = {Two's company, three's a crowd: a case study of crowdsourcing software development},
year = {2014},
isbn = {9781450327565},
publisher = {Association for Computing Machinery},
address = {New York, NY, USA},
doi = {10.1145/2568225.2568249},
booktitle = {Proceedings of the 36th International Conference on Software Engineering},
pages = {187–198},
numpages = {12},
location = {Hyderabad, India},
series = {ICSE 2014}
}

@article{lutters_warstories_2007,
    title = {Revealing actual documentation usage in software maintenance through war stories},
    journal = {Information and Software Technology},
    volume = {49},
    number = {6},
    pages = {576-587},
    year = {2007},
    note = {Qualitative Software Engineering Research},
    issn = {0950-5849},
    doi = {https://doi.org/10.1016/j.infsof.2007.02.013},
    author = {Wayne G. Lutters and Carolyn B. Seaman},
}

@article{seaman_qualitative_2025,
    title = {Qualitative {Research} {Methods} in {Software} {Engineering}: {Past}, {Present}, and {Future}},
    volume = {51},
    copyright = {https://ieeexplore.ieee.org/Xplorehelp/downloads/license-information/IEEE.html},
    issn = {0098-5589, 1939-3520, 2326-3881},
    shorttitle = {Qualitative {Research} {Methods} in {Software} {Engineering}},
    doi = {10.1109/TSE.2025.3538751},
    number = {3},
    urldate = {2025-08-29},
    journal = {IEEE Transactions on Software Engineering},
    author = {Seaman, Carolyn B. and Hoda, Rashina and Feldt, Robert},
    month = mar,
    year = {2025},
    pages = {783--788},
}

\end{document}